\documentstyle[epsfig,fleqn]{basi}
\begin{document}
\title[Optical observations of GRB 021004 afterglow ] 
{\bf Optical observations of the bright long duration peculiar GRB 021004 
afterglow }
\author[Pandey et al.]{S.B. Pandey$^1$, D.K. Sahu$^{2,3}$, L. Resmi$^{4,5}$,
R. Sagar$^{1,3}$, G.C. Anupama$^3$, 
\newauthor
D. Bhattacharya$^4$, V. Mohan$^1$, T.P. Prabhu$^3$, B.C. Bhatt$^{2,3}$, J.C. Pandey$^1$,
\newauthor
Padmaker Parihar$^{2,3}$ and A.J. Castro-Tirado$^{6}$ \\ \\
$^1$ State Observatory, Manora Peak, Nainital -- 263 129, India\\
$^2$ Center for Research \& Education in Science \& Technology, Hosakote, Bangalore -- 562 114, India\\
$^3$ Indian Institute of Astrophysics, Bangalore -- 560 034, India\\
$^4$ Raman Research Institute, Bangalore -- 560 080, India\\
$^5$ Joint Astronomy Programme, Indian Institute of Science, Bangalore -- 560 012, India\\
$^6$ Instituto de Astrof\'isica de Andaluc\'ia, P.O. Box 03004, 
   E-18080, Granada, Spain\\}
\pubyear{2003}
\volume{31}

\date{Received 2002 November 6; accepted 2003 February 10}

\maketitle
\begin{abstract}
The CCD magnitudes in Johnson $B,V$ and Cousins $R$ and $I$ photometric
passbands are determined for the bright long duration GRB 021004 afterglow from
2002 October 4 to 16 starting $\sim$ 3 hours after the $\gamma-$ray burst.
Light curves of the afterglow emission in $B$,$V$,$R$ and $I$ passbands are
obtained by combining these measurements with other published data. 
The earliest optical emission appears to originate in a revese shock.
Flux decay of the afterglow shows a very uncommon variation
relative to other well-observed GRBs. Rapid light variations, especially
during early times ($\Delta t < 2$~days) is superposed on an underlying
broken power law decay typical of a jetted afterglow. The flux decay
constants at early and late times derived from least square fits to the
light curve are $0.99\pm0.05$ and $2.0\pm0.2$ respectively, with a jet break
at around 7 day.  Comparison with a standard fireball model indicates a total
extinction of $E(B-V)=0.20$ mag in the direction of the burst. Our low-resolution 
spectra corrected for this extinction provide a spectral slope 
$\beta = 0.6\pm0.02$.  This value and the flux decay constants agree well with 
the electron energy index $p\sim 2.27$ used in the model.
The derived jet opening angle of about $7^{\circ}$ implies a total
emitted gamma-ray energy $E_{\gamma} = 3.5\times10^{50}$~erg at a cosmological
distance of about $20$~Gpc. Multiwavelength observations indicate association
of this GRB with a star forming region, supporting the case for collapsar
origin of long duration GRBs.
\end{abstract}


\section{Introduction}
In recent years, both photometric and spectroscopic optical observations  of 
Gamma-Ray Burst (GRB) afterglows have provided valuable information about the 
emission from GRBs. While spectral lines have been used to determine redshift 
distances and to study the host galaxies, photometric light curves have 
unravelled the physical parameters and dynamical evolution of GRB afterglows
(cf. Panaitescu \& Kumar 2002, Sagar 2001, 2002 and references therein). 

A long duration GRB 021004 ($\equiv$ H2380) triggered at 12$^h$06$^m$13.$^s$57 
UT on 4 October 2002 was detected by the HETE FREGATE, WXM, and soft $X-$ray 
camera (SXC) instruments (Shirasaki et al. 2002). The burst 
had a duration of $\sim$ 100 seconds in both FREGATE 8-40 Kev and WXM 2-25 Kev 
bands. Analyses of the FREGATE and WXM data by Lamb et al. (2002) show that the 
spectrum of the burst is well characterized by a single power-law with slope 
1.64$\pm$0.09. The burst fluences are 0.75, 1.8 and 3.2 $\mu$erg/cm$^2$ in the 
energy bands of 2 -- 25 Kev, 50 -- 300 Kev and 7 -- 400 Kev respectively. The 
fluence ratio S(2-25)/S(50-300) = 0.42 indicates that it is an $X-$ray rich GRB.

The SXC coordinates of the burst reported by Doty et al. (2002) are 
$\alpha = 00^h 26^m 55.^s75, \delta = +18^{\circ} 56^{'} 18.^{''}6$ (J2000). A 
relatively bright with $R \sim 15.5$ mag optical afterglow (OA) of the GRB 
021004 was discovered by Fox (2002) about 9 minutes after the burst at 
$\alpha = 00^h 26^m 54.^s687,  \delta = +18^{\circ} 55^{'} 41.^{''}3$ (J2000)
with an uncertainty of $0.^{''}5$ in each coordinates. The astrometric position 
of the OA determined by Henden \& Levine (2002) is
$\alpha = 00^h 26^m 54.^s674,  \delta = +18^{\circ} 55^{'} 41.^{''}59$ (J2000)
with $\sim$ 50 mas external error in each coordinates. This is in excellent
agreement with the coordinates  given by Fox (2002). Thus, GRB 021004 becomes 
second burst after GRB 990123 whose OA could be observed within few minutes of 
the trigger of the event. At the location of OA, Wood-Vasey et al. (2002) found
no source brighter than $R \sim 22$ mag on images taken on 3 October 2002 at 
$07^h 24^m 18^s, 07^h 54^m 50^s$ and $08^h 25^m 19^s$ UT while Sako \& Harrison
(2002) report a fading $X-$ray source with a power-law time slope of 
$-1.1\pm0.1$ using the Chandra $X-$ray observations taken with the High-Energy 
Transmission Grating on 2002 October 5, about 20.5 hours after the burst. Almost
within a day after the burst, the radio afterglow was also detected by Frail 
\& Berger (2002) at 22.5 GHz; by Pooley (2002) at 15 GHz and by Bremer \& 
Castro-Tirado (2002) at 86 GHz. The polarimetric observations taken on 
2002 October 05.077 and 08.225 UT indicate almost zero $V-$band intrinsic 
polarization for the OT (Covino et al. 2002; Rol et al. 2002, Wang et al. 2002).

Based on the detection of ionised Mg, Mn and Fe absorption features, Fox et al. 
(2002) indicated two redshift values, z = 1.38 and 1.60. 
Eracleous et al. (2002) and Anupama et al. (2002) also confirm the presence of 
two absorption systems. Chornock \& Filippenko (2002) identified in addition
to them, the emission lines at z = 2.323. The existence of this multi-component 
 z $\sim$ 2.3 redshift systems was also confirmed by Castro-Tirado et al. (2002),
Djorgovski et al. (2002), Mirabal et al. (2002b), Salamanca et al. (2002) and
Savaglio et al. (2002) in the high resolution spectrum of the OA.
The spectroscopic variability studied by Matheson et al. (2003) indicates that 
there is a colour evolution with the OA becoming redder with time, implying a 
(B-V) increase of about 0.2 - 0.3 mag over the first three days. The spectrum 
of the OA consists of a blue continuum with several absorption features  
corresponding to two intervening metal-line systems at $z =$ 1.380 and 1.602 
and one set of lines at a redshift of z = 2.323, apparently intrinsic to the 
host galaxy of the GRB. M$\o$ller et al. (2002), on the other hand, identify 
absorption lines from five systems at 
z = 1.3806, 1.6039, 2.2983, 2.3230 and 2.3292
along with an emission line at z = 2.3351.
 
There are no photometric standards in the field of GRB 021004 and the 
photometric calibrations published after the burst by Weidinger et al. (2002)
and Henden (2002) show a zero-point difference of 0.12 mag in $R$. A 
comparison of Henden (2002) photometry with that of Barsukova et al. (2002)
for 3 common star indicates that former is brighter by 0.2 to 0.3 mag in $B$; 
fainter by 0.01 to 0.11 mag in $V$ but agrees within errors in $R$. For 
reliable determination of the OA magnitudes, secured photometric calibrations 
are needed. In order to provide them, we imaged the field of GRB 021004 along 
with SA 92 standard region of Landolt (1992). A total of 40 secondary stars
in the field have been calibrated and their standard $BVRI$ magnitudes are given
here. Our observations started about 3 hour after 
the burst and are valuable for dense temporal coverage of the light curve. We 
present the details of our optical observations in the next section, and  
discuss the optical light curves and other results in the remaining sections.

\section { Optical observations, data reductions and calibrations }  

The broad band photometric and low-resolution spectroscopic optical observations
obtained for the GRB 021004 afterglow are described below along with their
data reduction and calibration. 
 
\subsection{Broad band photometric data}

The broad band Johnson $BV$ and Cousins $RI$ observations of the OA were carried
out between 4 to 16 October 2002 using 2-m Himalayan Chandra Telescope (HCT) of 
the Indian Astronomical Observatory (IAO), Hanle and the 104-cm Sampurnanand 
telescope of the State Observatory, Nainital. At Nainital, one pixel of the 
2048 $\times$ 2048 pixel$^{2}$ size CCD chip corresponds to 0.$^{''}$38 square, 
and 
the entire chip covers a field of $\sim 13^{\prime}\times 13^{\prime}$ on the 
sky. The gain and read out noise of the CCD camera are 10 $e^-/ADU$ and 5.3 
$e^-$ respectively.
At Hanle, one pixel corresponds to 0.$^{''}$3 square, and the entire chip 
covers a field of $\sim 10^{\prime} \times 10^{\prime}$ on the sky, it has a read
out noise of 4.95 $e^-$ and gain is 1.23 $e^-/ADU$. From Nainital, the 
CCD $BVRI$ observations of the OA field along with Landolt (1992) standard SA 
92 region were obtained on 13/14 October 2002 during good photometric sky 
conditions for photometric calibration. During the observing run, several 
twilight flat field and bias frames were also obtained for the CCD calibrations. 
 
The CCD frames were cleaned using standard procedures. Image processing was done
using ESO MIDAS, NOAO IRAF and DAOPHOT softwares. Atmospheric extinction 
coefficients determined from the Nainital observations of SA 92 bright stars are 0.34, 
0.22, 0.17 and 0.14 mag in $B, V, R$ and $I$ filters respectively on the night 
of 13/14 October 2002. They are used in our further analyses. There are nine 
standard stars in the SA 92 region. 
They cover a wide range in colour ($0.64 < (V-I) < 1.84$) as well as in 
brightness ($12.5 < V < 15.6$). The transformation coefficients were determined 
by fitting least square linear regressions to the following equations.
 
\noindent $b_{CCD} = B - (0.036\pm0.01) (B-V) + (5.08\pm0.02) $  \\ 
	$v_{CCD} = V - (0.051\pm0.01) (B-V) + (4.56\pm0.01) $  \\ 
	$r_{CCD} = R -(0.003\pm0.01) (V-R) + (4.38\pm0.01) $  \\ 
	$i_{CCD} = I - (0.026\pm0.01) (V-R)  + (4.87\pm0.02) $  \\ 
where $BVRI$ are standard magnitudes and $v_{CCD}, b_{CCD}, r_{CCD} $ and 
$i_{CCD}$ represent the instrumental magnitudes normalized for 1 second of 
exposure time and corrected for atmospheric extinction. The errors in the colour
coefficients and zero points are obtained from the deviation of data points 
from the linear relation. Using these transformations, $BVRI$ photometric 
magnitudes of 40 secondary standard stars are determined in the GRB 021004 
field and their average values are listed in Table 1. The $(X,Y)$ CCD pixel 
coordinates are converted into $\alpha_{2000}, \delta_{2000}$ values using the 
astrometric positions given by Henden (2002). All these stars have
been observed 3 to 17 times in a filter and have internal photometric accuracy  
better than 0.01 mag. Henden (2002) also provides the $UBVRI$ photometry for a 
large number of stars in the field. A comparison in the 
sense present minus Henden (2002) value yields small systematic zero-point 
differences of $-0.002\pm0.03, 0.013\pm0.02, 0.02\pm0.026$ and $0.03\pm0.03$ mag
in $B, V, R$ and $I$ filters respectively. These numbers are based on 25 common 
stars having range in brightness from $V = 14$ to 18 mag and can be accounted 
for in terms of zero-point errors in the two photometries. There is no colour 
dependence in the photometric differences. We therefore conclude that 
photometric calibration used in this work is secure.

Several short exposures up to a maximum of 15 minutes were generally given
while imaging the OA (see Table 2). In order to improve the signal-to-noise 
ratio of the OA, the data have been binned in $2 \times 2$ pixel$^2$ and also 
several bias corrected and flat-fielded CCD images of OA field taken on a night 
are co-added in the same filter, when found necessary. From these images, 
profile-fitting magnitudes are determined using DAOPHOT software. For
determining the difference between aperture and profile fitting magnitudes,
we constructed an aperture growth curve of the well isolated stars and used
them to determine aperture (about 5 arcsec) for the magnitudes of the OA. 
They are calibrated differentially with
respect to the secondary standards listed in Table 1 and the values 
derived in this way are given in Table 2. They supersede the values published 
earlier by Sahu et al. (2002).

The secondary standards are also used for calibrating other photometric 
measurements of OA published by the time of paper submission by 
Bersier et al. (2003), Covino et al. (2002), Di Paola et al. (2002), Fox (2002), Garnavich 
\& Quinn (2002), Holland et al. (2003), Halpern et al. (2002a, b), Malesani 
et al. (2002a, b), Masetti et al. (2002), Matsumoto et al. (2002), Mirabal et 
al. (2002a, b), Oksanen et al. (2002), Rhoads et al. (2002), 
 Stefanon et al. (2002), Williams et al. (2002),
Winn et al. (2002) and Zharikov et al. (2002). In order to avoid errors 
arising due to different photometric calibrations, we have used only those 
published $BVRI$ photometric measurements whose magnitudes could be determined 
relative to the stars given in Table 1. The $JHK$ magnitudes are adopted from 
Di Paola et al. (2002), Rhoads et al. (2002) and Stefanon et al. (2002). 
The distribution of photometric data points taken from the literature and from 
the present measurements are $N(U,B,V,R,I,J,H,K) = (6,25,31,197,23,2,3,3)$ and
$N(B,V,R,I) = (15,27,67,29)$ respectively. Thus, a total of 428 photometric 
data points in eight passbands are there for our analysis in the optical and
near-IR region. 

\begin{table}  
{\bf Table 1.}~The identification number(ID), $(\alpha , \delta)$ for epoch 2000, 
standard $V, (B-V), (V-R)$ and $(R-I)$ photometric magnitudes of the stars in 
the GRB 021004 region are given. N(B,V,R,I) denotes the number of observations 
taken in $B, V, R$ and $I$ filters respectively. Star 23 is the comparison star mentioned by 
Henden (2002). 
 
\begin{center} 
\scriptsize
\begin{tabular}{ccc cc ccl} \hline  
ID & $\alpha_{2000}$ & $\delta_{2000}$ & $V$& $B-V$ & $V-R$ & $V-I$ & 
N(B,V,R,I) \\ 
    & (h m s) & (deg m s) & (mag) & (mag) & (mag) & (mag) & \\ \hline 
  1&00 26 29&18 54 28& 17.287&  0.750&  0.428&  0.765& (9,9,17,9)\\
  2&00 26 31&18 55 20& 14.412&  0.876&  0.487&  0.873& (9,10,16,10)\\
  3&00 26 32&18 55 56& 17.321&  0.813&  0.467&  0.908& (9,10,17,9)\\
  4&00 26 32&18 57 09& 15.708&  1.097&  0.637&  1.151& (9,10,17,9)\\
  5&00 26 34&18 54 43& 16.924& 0.642&  0.379&  0.703& (9,10,17,9)\\
  6&00 26 34&18 57 00& 17.701&  0.703&  0.383&  0.746& (9,10,17,10)\\
  7&00 26 35&18 51 49& 15.743&  1.043&  0.621&  1.090& (9,10,17,9)\\
  8&00 26 37&18 54 51& 14.399&  0.653&  0.373&  0.706& (9,10,16,10)\\
  9&00 26 38&18 58 19&14.704& 0.787&  0.423&  0.802& (9,10,16,10)\\
 10&00 26 39&18 56 01& 13.234&  1.072&  0.557&  1.044& (9,9,7,3)\\
 11&00 26 44&18 51 56& 14.199& 0.669&  0.372&  0.683& (9,10,16,10)\\
 12&00 26 46&18 55 24& 13.906& 0.717&  0.385&  0.767& (9,10,16,10)\\
 13&00 26 48&18 56 33& 16.746&  0.900&  0.504&  1.016& (9,10,17,10)\\
 14&00 26 51&18 54 37& 17.514&  0.606&  0.340&  0.694& (9,10,17,10)\\
 15&00 26 51&18 59 10& 17.469&  1.481&  1.111&  2.426& (9,10,17,10)\\
 16&00 26 51&18 57 47& 17.862&  0.760&  0.429&  0.887& (9,10,17,10)\\
 17&00 26 52&18 55 12& 14.449&  1.072&  0.609&  1.111&(9,10,17,9) \\
 18&00 26 53&19 02 23& 15.646&  0.885&  0.506&  0.942& (9,10,17,10)\\
 19&00 26 54&18 52 24& 16.125&  0.740&  0.413&  0.774& (9,9,17,9)\\
 20&00 26 54&18 53 45& 17.985&  0.580&  0.349&  0.668& (8,9,15,9)\\
 21&00 26 58&18 56 08&16.717& 0.617&  0.355&  0.725& (9,10,17,10) \\
 22&00 26 58&18 59 51& 11.670&  0.640&  0.386&  0.758& (3,4,3,3) \\
 23&00 26 59&18 56 57& 16.273& 1.149&  0.711&  1.360& (9,10,17,10)\\
 24&00 27 01&18 51 16& 15.680& 1.075& 0.638&  1.142& (9,10,17,10) \\
 25&00 27 01&18 54 16& 17.325&  0.484&  0.309&  0.666& (9,10,17,9) \\
 26&00 27 05&18 55 50& 15.352&  0.618&  0.359&  0.726& (9,10,17,10) \\
 27&00 27 05&18 55 51& 17.356&  0.652&  0.394&  0.810& (9,10,17,10) \\
 28&00 27 06&18 58 18& 16.139& 0.626& 0.366& 0.744& (9,10,17,10) \\
 29&00 27 06&19 03 10& 17.506&  1.054&  0.652&  1.174& (7,7,11,5) \\
 30&00 27 07&18 52 21& 17.958&  0.662&  0.349&  0.735& (9,9,17,10) \\
 31&00 27 08&18 57 08& 17.118&  0.777&  0.449&  0.878& (9,10,17,10) \\
 32&00 27 08&18 59 36& 13.580& 0.912& 0.510&  0.961& (9,10,16,10)\\
 33&00 27 09&19 01 59& 16.041&  1.151&  0.690&  1.226& (9,10,17,10) \\
 34&00 27 10&19 00 48& 17.769&  1.463&  0.994&  2.024& (9,10,17,10) \\
 35&00 27 12&18 55 39& 16.722&  1.139&  0.716&  1.297& (9,10,17,10)  \\
 36&00 27 13&18 51 35& 17.476&  0.842&  0.496&  0.946& (9,9,17,10)   \\
 37&00 27 13&18 54 08& 16.125& 0.914& 0.573&  1.068& (9,10,17,10) \\
 38&00 27 13&19 02 45& 17.420&  1.427&  0.916&  1.686& (8,9,15,9)  \\
 39&00 27 14&18 56 24& 16.096&  0.832&  0.480&  0.938& (9,8,14,9)  \\
 40&00 27 15&18 59 12& 15.252&1.055&  0.610& 1.145& (5,6,14,8) \\
\hline
\end{tabular} 
\end{center} 
\end{table} 

\normalsize
\begin{table}
 {\bf Table 2.}~CCD BVRI broad band optical photometric observations of 
the GRB 021004 afterglow. At Hanle, 2-m HCT was 
used while at Nainital, 104-cm Sampurnanand optical telescope was used. 

\begin{center}
\scriptsize
\begin{tabular}{ccll|ccll} \hline 
Date (UT) of & Magnitude & Exposure time & Telescope& Date (UT) of & Magnitude &
Exposure time & Telescope  \\
2002 October& (mag)&(Seconds)& &2002 October&(mag)&(Seconds)& \\   \hline 
\multicolumn{3}{c}{\bf $B-$ passband}&&\multicolumn{3}{c}{\bf $V-$ passband (continued)}&\\
04.725000 &18.37$\pm$0.01&600 &104-cm&07.878831&20.80$\pm$0.02&3$\times$600 + 400&HCT \\
04.728681&18.43$\pm$0.01&600 & HCT & 08.818565&21.20$\pm$0.03&2$\times$600&HCT \\ 
04.733333&18.43$\pm$0.01&600 &104-cm&08.856076&21.21$\pm$0.03&2$\times$600&HCT\\ 
04.741667&18.52$\pm$0.01&600 &104-cm&10.778611&21.77$\pm$0.04&3$\times$600&HCT \\ 
04.833333&18.87$\pm$0.01&600 &104-cm&10.828819&21.66$\pm$0.03&3$\times$600&HCT \\ 
04.834109&18.88$\pm$0.02&600 & HCT&10.881933&21.71$\pm$0.04&3$\times$600&HCT \\ 
04.841667&18.95$\pm$0.01&600 &104-cm&11.851632&21.91$\pm$0.03&4$\times$600&HCT \\ 
04.850000&18.96$\pm$0.01&600 &104-cm& \multicolumn{3}{c}{\bf $I-$ passband}&\\
04.928472&19.19$\pm$0.02&300 &104-cm&04.671528&16.48$\pm$0.01&300&104-cm \\ 
04.929329&19.21$\pm$0.02&600 & HCT&04.679861&16.56$\pm$0.01&300&104-cm   \\ 
04.960544&19.31$\pm$0.02&600 & HCT&04.684722&16.58$\pm$0.01&300&104-cm   \\ 
04.982639&19.33$\pm$0.02&300 &104-cm&04.745602&17.20$\pm$0.01& 300& HCT \\ 
05.673588&20.33$\pm$0.02&1200& HCT&04.768750&17.31$\pm$0.01&300&104-cm   \\ 
05.809676&20.53$\pm$0.02&1200& HCT&04.773611&17.36$\pm$0.01&300&104-cm   \\ 
05.871713&20.62$\pm$0.02&1200& HCT&04.778472&17.32$\pm$0.01&300&104-cm   \\ 
 \multicolumn{3}{c}{\bf $V-$ passband}&&04.850660&17.54$\pm$0.01&300&HCT \\ 
04.691667&17.53$\pm$0.01&600& 104-cm&04.861806&17.48$\pm$0.02&100&104-cm    \\ 
04.700000&17.64$\pm$0.01&600& 104-cm&04.936806&17.68$\pm$0.04&50 &104-cm    \\ 
04.713889&17.70$\pm$0.01&600& 104-cm&04.938032&17.77$\pm$0.02&300&HCT    \\ 
04.737870&17.99$\pm$0.02&400& HCT &04.976389&17.84$\pm$0.06&50 &104-cm      \\ 
04.784722&18.29$\pm$0.01&600& 104-cm&05.646030&18.83$\pm$0.02&600& HCT    \\ 
04.792361&18.31$\pm$0.01&500& 104-cm&05.712882&18.86$\pm$0.01&600& HCT    \\ 
04.800000&18.33$\pm$0.01&500& 104-cm&05.834213&19.11$\pm$0.02&600& HCT    \\ 
04.843495&18.45$\pm$0.02&400& HCT &05.895949&19.19$\pm$0.02&600& HCT      \\ 
04.856250&18.45$\pm$0.01&200& 104-cm&06.756343&19.51$\pm$0.02&600 + 300& HCT    \\ 
04.932640&18.70$\pm$0.01&100& 104-cm&06.775690&19.48$\pm$0.02&600 + 300& HCT    \\ 
04.944792&18.73$\pm$0.02&400& HCT & 06.893345&19.62$\pm$0.03&2$\times$500& HCT     \\ 
04.965278&18.83$\pm$0.01&100& 104-cm& 07.908241&20.05$\pm$0.05&4$\times$250 + 300& HCT\\ 
04.969132&18.84$\pm$0.02&400&HCT&07.927083&19.85$\pm$0.11&6$\times$300&104-cm     \\ 
05.658657&19.81$\pm$0.02&900& HCT &08.675694&20.31$\pm$0.07&6$\times$300&104-cm    \\ 
05.680671&19.86$\pm$0.01&3$\times$900&104-cm&08.869317&20.35$\pm$0.05& 400 + 300& HCT  \\ 
05.725729&19.89$\pm$0.02&600& HCT &08.884282&20.35$\pm$0.04& 3$\times$300& HCT \\ 
05.823009&20.02$\pm$0.02&600& HCT&08.903009&20.45$\pm$0.05& 3$\times$300& HCT \\ 
05.886007&20.13$\pm$0.03&600& HCT&09.829167&20.46$\pm$0.10&2$\times$900&104-cm \\ 
06.730463&20.48$\pm$0.02&2$\times$600&HCT& 10.809030&21.00$\pm$0.14&2$\times$900&104-cm\\ 
06.864549&20.54$\pm$0.02&3$\times$600&HCT&11.760470&21.30$\pm$0.14&4$\times$300&104-cm \\ 
&             &            &             &14.750690&21.64$\pm$0.18&3$\times$900&104-cm\\ 
\hline
\end{tabular} 
\end{center} 
\end{table} 
\normalsize

\normalsize
\begin{table}
 {\bf Table 2.}~(Continued)

\begin{center}
\scriptsize
\begin{tabular}{ccll|ccll} \hline 
Date (UT) of & Magnitude & Exposure time & Telescope& Date (UT) of & Magnitude &
Exposure time & Telescope  \\
2002 October& (mag)&(Seconds)& &2002 October&(mag)&(Seconds)& \\   \hline 
\multicolumn{3}{c}{\bf $R-$ passband}&&\multicolumn{3}{c}{\bf $R-$ passband (continued)}&\\
04.625694&16.74$\pm$0.01&300& 104-cm&05.886806&19.73$\pm$0.02&3$\times$900& 104-cm \\
04.631944&16.73$\pm$0.01& 300&104-cm  &05.908704&19.70$\pm$0.02&900&HCT \\ 
04.637500 &16.75$\pm$0.01&300& 104-cm&05.952072&19.72$\pm$0.02& 900& HCT\\ 
04.642361 &16.81$\pm$0.01& 300 & 104-cm &06.648611&20.04$\pm$0.02&4$\times$900&104-cm \\ 
04.654861 &16.87$\pm$0.01& 300 & 104-cm&06.700972&19.98$\pm$0.02&3$\times$600& HCT  \\ 
04.684051 &17.06$\pm$0.01& 60 & HCT&06.807998&20.08$\pm$0.02&3$\times$600& HCT \\ 
04.690694 &17.12$\pm$0.01& 300 & HCT &06.835880&20.09$\pm$0.01&3$\times$600& HCT \\ 
04.697257 &17.22$\pm$0.01& 300 & HCT& 06.926921&20.15$\pm$0.02&3$\times$600& HCT\\
04.703125 &17.27$\pm$0.01& 300 & HCT&07.820856&20.38$\pm$0.02&4$\times$400&HCT \\ 
04.709213 &17.28$\pm$0.02& 300 & HCT&07.887847&20.37$\pm$0.02&2$\times$400 + 500& HCT   \\ 
04.715000 &17.31$\pm$0.02& 300 & HCT&07.898611&20.50$\pm$0.09&5$\times$300&104-cm   \\ 
04.720972 &17.41$\pm$0.02& 300 & HCT&07.948611&20.43$\pm$0.13&3$\times$300&104-cm \\ 
04.749306 &17.73$\pm$0.01& 300 & 104-cm&08.645139&20.69$\pm$0.04&6$\times$300 & 104-cm   \\ 
04.753438 &17.76$\pm$0.02& 300 & HCT&08.769491&20.71$\pm$0.03& 600 + 500& HCT   \\ 
04.754167 &17.74$\pm$0.01& 300 & 104-cm&08.793299&20.72$\pm$0.03&2$\times$600&HCT   \\ 
04.759028 &17.77$\pm$0.01& 300 & 104-cm&09.727234&20.96$\pm$0.03&3$\times$500&HCT    \\ 
04.763194 &17.81$\pm$0.01& 300 & 104-cm&09.766319&20.99$\pm$0.02&3$\times$600&HCT    \\ 
04.825417 &17.93$\pm$0.01& 300 & HCT&09.800000&21.12$\pm$0.06&3$\times$900 & 104-cm     \\ 
04.856667 &18.07$\pm$0.01& 300 & HCT &09.810069&20.95$\pm$0.02&4$\times$600&HCT      \\ 
04.859028 &18.06$\pm$0.02& 100 & 104-cm&10.706250&21.29$\pm$0.06&2$\times$900 & 104-cm  \\ 
04.920810 &18.29$\pm$0.02& 300 & HCT&10.711505&21.23$\pm$0.03&2$\times$500&HCT    \\ 
04.934722 &18.35$\pm$0.04& 50 & 104-cm&10.732951&21.22$\pm$0.03& 400 + 300& HCT   \\ 
04.940278 &18.31$\pm$0.02& 300 & 104-cm &11.722220&21.55$\pm$0.11&2$\times$900 + 2$\times$300 & 104-cm   \\ 
04.945139 &18.38$\pm$0.02& 300 & 104-cm&11.799398&21.44$\pm$0.03&3$\times$500& HCT    \\ 
04.954167 &18.37$\pm$0.02& 300 & 104-cm &11.846667&21.43$\pm$0.03&3$\times$500& HCT    \\ 
04.967361 &18.48$\pm$0.03& 50 & 104-cm & 13.675613&21.91$\pm$0.05&5$\times$500& HCT      \\ 
05.616771&19.28$\pm$0.02&900& HCT & 13.716991&21.93$\pm$0.05&6$\times$400& HCT\\ 
05.633576&19.34$\pm$0.02&900& HCT&13.756940&21.71$\pm$0.08&2$\times$900 & 104-cm     \\ 
05.641667&19.42$\pm$0.02&4$\times$600&104-cm &14.659720&21.87$\pm$0.11&2$\times$900 + 1800 & 104-cm  \\ 
05.703252&19.42$\pm$0.01&600& HCT & 14.773877&22.04$\pm$0.07&5$\times$500& HCT\\ 
05.736042&19.46$\pm$0.02&600&HCT&14.815926&22.09$\pm$0.06&6$\times$500& HCT \\ 
05.785718&19.51$\pm$0.01&300& HCT&15.847570&22.22$\pm$0.07&4$\times$600& HCT  \\ 
05.793356&19.55$\pm$0.01&600& HCT & 15.852780&22.48$\pm$0.52&1$\times$1800& 104-cm \\ 
05.846794&19.59$\pm$0.01&600& HCT & 16.688738&22.46$\pm$0.15&10$\times$600& HCT \\ 
05.855926&19.62$\pm$0.02&600& HCT &&&& \\ 
\hline
\end{tabular} 
\end{center} 
\end{table} 
\subsection {Spectroscopic observations}

CCD low-resolution spectra of the OA were obtained, from IAO, on 2002 October 4.789,
4.806, 4.876 and 4.894 UT, using the Hanle Faint Object Spectrograph Camera
instrument. The epochs correspond to 0.285, 0.302, 0.372 and 0.39 day 
respectively after the burst. The exposure times were 900s for the first two 
and 1200s for the last two spectra. They were obtained at a resolution of 
18~\AA, using a slit width of 2${''}$, covering a wavelength range of
5200--9000~\AA. Spectrophotometric standard BD+28$^{\circ}$~4211 was observed 
with a wider slit of 15${''}$ width. 

All spectra were bias subtracted, flat-field corrected, extracted and wavelength
calibrated in the standard manner using the IRAF reduction package. The spectra
were corrected for instrumental response and brought to a flux scale using the
spectrophotometric standard. Since the position angle of the slit was not
along the parallactic angle (Filippenko 1982), and the observations were made
at an airmass $\sim 2.2$, the fluxes of the OA have been calibrated using 
zero points derived from $BVRI$ photometry. The spectra have been corrected for
a total (Galactic and/or host galaxy) extinction of $E(B-V)=0.20$ mag (see
section 4 for details) and shown in Fig 1. The spectrum shows 
a blue continuum with superposed absorption features. The absorption systems are
identified with two intervening metal-line systems at z = 1.38 and 1.60. The 
line center of the absorption features, their identification and the inferred 
redshift values are listed in Table 3. The line systems are marked in Fig 1.
Present results supercede the analysis presented by Anupama et al. (2002) and
agree well with other spectroscopic determinations published in the literature.

A single power law $F_\nu \propto \nu^{-\beta}$ was found to fit continuum
of the observed spectra. A chi-squared minimization for the power law yields 
an index of $\beta = 0.59\pm 0.02$ for the averaged, $E(B-V) = 0.20$ mag
extinction corrected spectrum. This value becomes
1.07$\pm$0.06, if the spectra is corrected only for the Galactic extinction
with $E(B-V) = 0.06$ mag which then agrees fairly well with the value of 
$\beta = 0.96\pm 0.03$ derived by Matheson et al. (2003).

\begin{table}
{\bf Table 3.}~Absorption lines in GRB 021004 afterglow spectrum.

\begin{center} 
\begin{tabular}{cccc} \hline
 Identification & \multicolumn {2}{c}{$\lambda$ in (\AA)} &  Redshift\\
    & Observed & Rest        & \\ \hline
 Fe II & 5577.8 &2343.5 & 1.380\\
 Fe II & 5649.3 &2373.7 & 1.380\\
 Fe II & 5669.2 &2382.0 & 1.380\\
 Fe II & 6185.0 &2599.0 & 1.379\\
 Fe II & 6194.6 &2382.0 & 1.601\\
 Mg II & 6655.3 &2795.5 & 1.381\\
 Mg II & 6672.7 &2802.7 & 1.381\\
 Fe II & 6733.1 &2585.9 & 1.604\\
 Fe II & 6762.7 &2599.4 & 1.602\\
 Mg II & 7273.1 &2795.5 & 1.602\\
 Mg II & 7292.9 &2802.7 & 1.602\\ \hline
\end{tabular}
\end{center}
\end{table}

\begin{figure}[t]
\begin{center}
{\epsfig{file=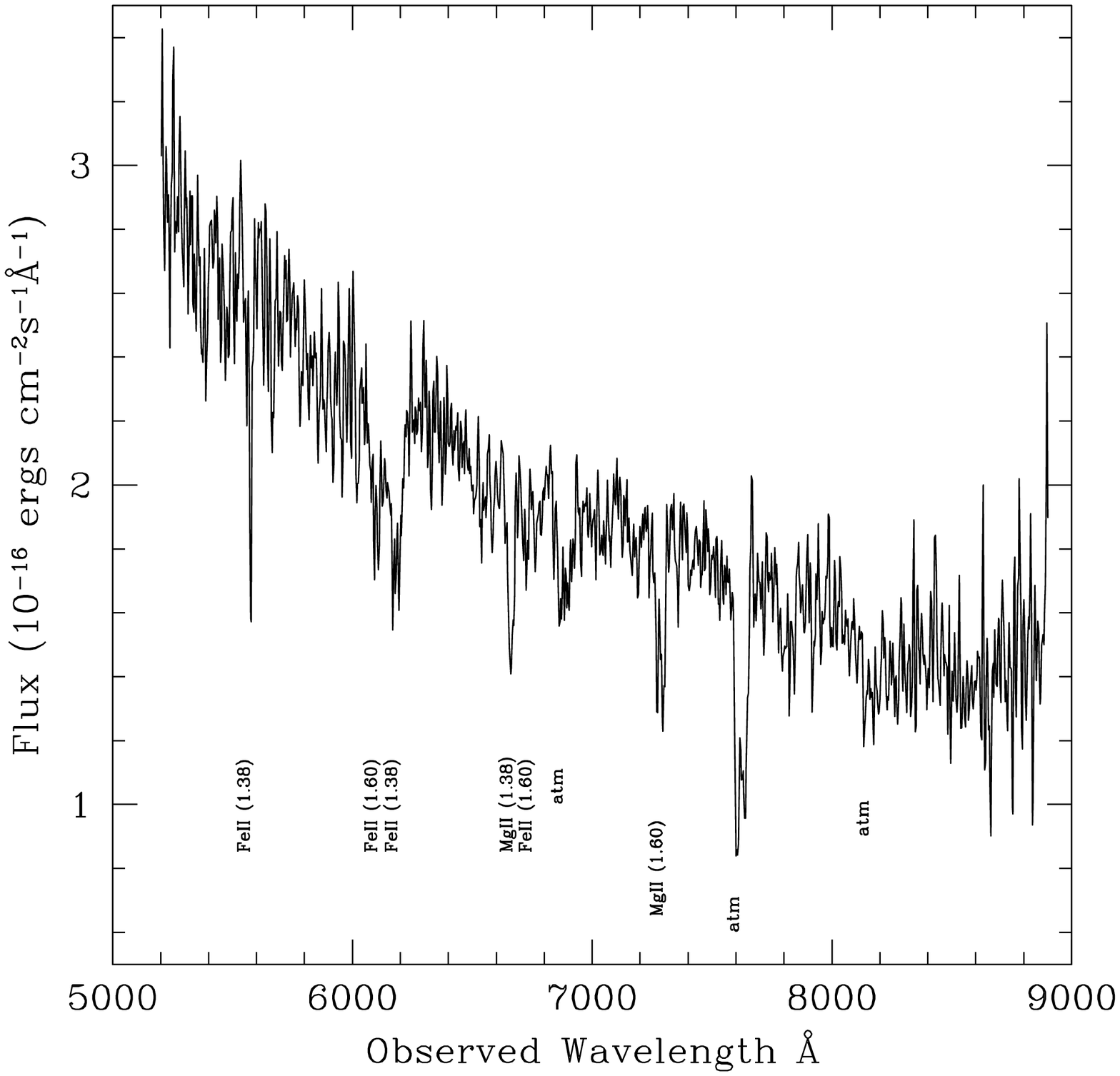,height=9cm,width=12cm}}
\end{center}\vspace*{-0.1cm}
\caption{\label{ospec} Optical spectrum of the GRB 0021004 OA corrected for
$E(B-V) = 0.20$ mag in the wavelength range 5500--9000~\AA. The absorption 
lines are marked along with the estimated redshift value.}
\end{figure}

\section{ Optical photometric light curves}

We have used the present measurements in combination with the published data to 
study the flux decay of GRB 021004 afterglow. Fig. 2 shows the plot of 
photometric measurements as a function of time. The X-axis is log ($\Delta t = 
t-t_0$) where $t$ is the time of observation and $t_0=$~2002 October 4.504325 UT
is the burst epoch.  All times are measured in unit of day. 

\begin{figure}[t]
\begin{center}
{\epsfig{file=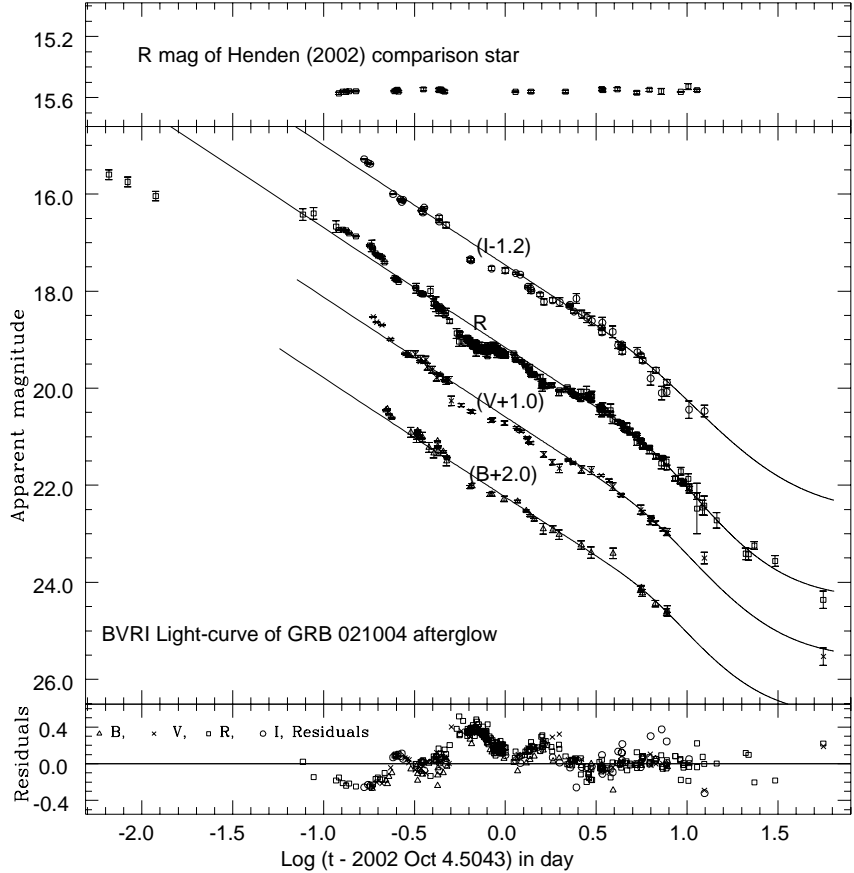,height=12cm}}
\end{center}\vspace*{-0.8cm}
\caption{\label{light} Light curves of GRB 0021004 afterglow in optical 
photometric $BVRI$ passbands are shown in the middle panel.
Marked vertical offsets have been applied to avoid overlapping of data 
points of different passbands. For comparison, R magnitude of Henden (2002) comparison star
is also plotted in the upper panel. The $BVRI$ band residuals in the sense observed
minus power-law fitted magnitudes are displayed in the lower panel.  }
\end{figure}

\subsection {Rapid variability in the $BVRI$ optical light curve}

The flux decay of most of the earlier GRB afterglows is generally well 
characterized by a single power law $F(t) \propto (t-t_0)^{-\alpha}$, where 
$F(t)$ is the flux of the afterglow at time $t$ and $\alpha$ is the decay 
constant. However, optical light curves of GRB 021004 (Fig. 2) show rapid 
variations with an overall flux decay
especially during $\Delta t < 2$ day. Among equally well monitored GRB OA, GRB 
021004 appears therefore peculiar. In order to see whether variability is 
correlated in $B, V, R$ and $I$ passbands or not, we derived photometric colours
using optical and near-IR data and list them in Table 4. Where necessary, 
measurements were interpolated between adjacent data points at one wavelength in
order to determine a contemporaneous value with another wavelength.
There is no evidence for statistically significant large variation in the photometric 
colours on these time scales. This  result is therefore contrary to the 
variability of spectroscopic colour reported by Matheson et al. (2003).

 In order to ensure that observed 
variability in the OA is not due to errors in photometric measurements, we also plotted
in Fig. 2, the R value of Henden's (2002) comparison star against time 
which showed no variability while Fig. 2 clearly indicates peculiar behaviour of the 
light curve showing achromatic variability in $BVRI$ passbands during early
phase. A large fraction of these observations have been carried out using the 1-m 
class optical telescopes. This indicates that in future large amount of observing time 
is available on these telescopes (cf. Sagar 2000), will  play
an important role in understanding the origin of such short term variability in 
the light curves of GRBs during early times. 

In order to analyse the rapid flux variations, we plot in the
bottom panel of Fig. 2, residuals after subtracting the best fitted power 
law values from the corresponding observed ones against time.
The variations appear to be achromatic and are clearly visible due to dense temporal
coverage of photometric observations. They have
a mean of 0.02$\pm$0.12, 0.07$\pm$0.15, 0.13$\pm$0.18 and 0.05$\pm$0.15 mag
in $B$,$V$,$R$ and $I$ filters respectively. We obtain a rough estimate of 
these time variations by fitting gaussian, which gives the FWHM values to 
be $\sim$ 11.5  and 21 hour for the bumps between $\Delta t$ = 0.25
to 1 day and 1.1 to 2.1 day respectively. These periods are considerably larger 
than 0.7 hour period found by Holland et al. (2002) and Jakobsson et al. (2003) 
for GRB 011211.
The peculiar nature of the GRB 021004 OA can be explained in terms of 
variable external density of the medium or variation in energy
of the blast wave with time (Nakar, Piran \& Granot 2003).
Lazzati el al. (2002) have also explained this peculiarity in terms of 
density enhancements of the surrounding medium.

\begin{table}
 {\bf Table 4.}~Broad band photometric colours of GRB 021004 OA at selected 
epochs. 
\begin{center}
\begin{tabular}{ccc ccc} \hline 
$\Delta t$&$(B-V)$ & $(V-I)$  & $(B-J)$ & $(B-H)$ & $(B-K)$ \\
 (in days)&  (mag) &(mag) &(mag) &(mag) &(mag)  \\ \hline
~0.22& 0.52$\pm$0.03 &  \\
~0.34& 0.53$\pm$0.03 & 1.05$\pm$0.03  \\
~0.46& 0.51$\pm$0.03 & 1.03$\pm$0.03  \\
~1.17& 0.49$\pm$0.03 & 0.99$\pm$0.04  \\
~1.37&0.52$\pm$0.04 & 0.95$\pm$0.06 & 2.2$\pm$0.15 & & 4.2$\pm$0.3\\
~1.66&0.57$\pm$0.05 & 0.93$\pm$0.08 & 2.3$\pm$0.15 & & 3.7$\pm$0.3\\
~2.01&0.60$\pm$0.06 & 1.00$\pm$0.06 &  & 3.3$\pm$0.15 &   \\
~2.98&0.65$\pm$0.10 & 0.98$\pm$0.10 &  & 3.2$\pm$0.15 &   \\
~5.57&0.63$\pm$0.10 & 1.01$\pm$0.10 &  & 3.4$\pm$0.15 &   \\
~7.73&0.58$\pm$0.10 & 0.75$\pm$0.20  \\
12.46&0.58$\pm$0.10 & 0.83$\pm$0.15  \\ \hline
\end{tabular}
\end{center}
\end{table}

\subsection {Parameters from optical light curves}

In the light curves there is steepening after $\Delta t$ $>$ 6 day appears to 
be present in $B$,$V$,$R$ and $I$ passbands. Achromatic fluctuations are also 
clear in $BVRI$ passbands.
In order to determine the flux decay and related parameters from the optical 
light curves following analyses have been carried out.

\begin{enumerate}
\item The earliest 3 data points in $R$ passband show a flux decay with $\alpha 
= 0.69\pm0.05$. This could be due to reverse shock emissions as the value of 
$\alpha$ is generally around 1.0 during early time flux decay of forward shock 
emissions (cf. Kobayashi \& Zhang 2003 and section 4). 
\item It is unlikely that emission from reverse shock will contribute 
significantly after $\Delta t > 0.1$ day and the steepening in the light curve
appears to be 6 days after the burst. We have therefore determined flux decay
constant for the OA using least square linear fit to the data points of $\Delta 
t < 5$ day and found values of 0.92$\pm$0.08, 0.93$\pm$0.03 and
1.13$\pm$0.04 in $V$,$R$ and $I$ passbands respectively. Average value of 
the early time flux decay constant of the OA is therefore 0.99$\pm$0.05. This 
is in good agreement with the values of early time flux decay constants of 
well observed GRB OA and also is as expected theoretically.

\item To determine the late time flux decay constant and break time, we fitted
the following empirical function (see Rhoads \& Fruchter 2000) which represents 
a broken power-law in the light curve in presence of underlying host galaxy.
\[
m=m_b + \frac{2.5}{s} [log_{10} \{ (t/t_b)^{\alpha_1s}
+(t/t_b)^{\alpha_2s}\} - log_{10}(2)] + m_g 
\]
where $\alpha_1$ and $\alpha_2$ are asymptotic power-law slopes at early and 
late times with $\alpha_1 < \alpha_2$ and $s > 0$ controls the sharpness of the 
break, with larger $s$ implying a sharper break. $m_b$ is the magnitude at the 
cross-over time $t_b$. $m_g$ is the magnitude of underlying host galaxy.
The function describes a light curve falling as 
$t^{-\alpha_1}$ at $t << t_b$ and $t^{-\alpha_2}$ at $t >> t_b$. 
In jet models, an achromatic break in the light curve is expected when the jet 
makes the 
transition to sideways expansion after the relativistic Lorentz factor
drops below the inverse of the opening angle of the initial beam. 
As there are rapid variations around overall early time flux
decay, we fitted the above function in $BVRI$ bands, for $\Delta t >$ 2 day to determine 
the parameters of the jet model. In order to 
avoid a fairly wide range of model parameters for a comparable $\chi^2$ due to degeneracy 
between $t_b, m_b, m_g, \alpha_1, \alpha_2$ and $s$, we have used fixed values of 
$\alpha_1 = 0.99$ and $s$ in our analyses and find that the minimum 
value of $\chi^2$ is achieved around $s = 4$. We also fixed the value of
$m_g$ for a minimum value of $\chi^2$ for different filters. 
The fitted values of host galaxy contributions $m_g$ are $\sim$ 24.79, 24.59, 24.35 
and 23.79 mag for $B$,$V$,$R$ and $I$ passbands respectively.
The least square fit values of the parameters 
$t_b$, $m_b$, and $\alpha_2$ are 6.51$\pm$0.12~day, 21.37$\pm$0.03 mag, and 
2.06$\pm$0.05 respectively in $R$ band, with a corresponding $\chi^2$ of 2.44 per 
degree of freedom ($DOF$). For $V$ passband fitted values of $t_b$, $m_b$, and $\alpha_2$ are
6.44$\pm$0.28~day, 21.79$\pm$0.06 mag, and 1.96$\pm$0.17 respectively with
$\chi^2$ 3.16 per $DOF$. For $B$ and $I$ filters we also fixed the value
of $t_b$ at 6.5~day to determine the values of $\alpha_2$. The values of $\alpha_2$ 
are 2.07$\pm$0.40 and 1.78$\pm$0.15 respectively for $B$ and $I$ filters
with $m_b$ values of 22.46$\pm$0.03 and 20.87$\pm$0.01 mag. For $B$ and $I$
filters $\chi^2$ values are 1.05 and 2.73 per $DOF$.
This indicates that the observed break in the light curve is sharp, unlike the smooth 
break observed in the optical light curve of GRB 990510 (cf. Stanek et al. 1999; Harrison et al. 
1999) but similar to the sharp break observed in the optical light curves of GRB
000301c (cf. Berger et al. 2000, Sagar et al. 2000, Pandey et al. 2001); GRB 000926 (cf. Harrison 
et al. 2001, Sagar et al. 2001a, Pandey et al. 2001); GRB 010222 (cf. Masetti et al. 2001; Sagar et 
al. 2001b; Stanek et al. 2001; Cowsik et al. 2001) and GRB 011211 (cf. 
Jakobsson et al. 2003). 
In Fig. 2 the best fit light curves obtained in this way for $BVRI$
passbands are shown. It can also be seen that our own 
observations follow the fitted curves very well and fill gaps in the published data. 
\end{enumerate}

In the light of above, we conclude that the parameters derived from the optical
$BVRI$ light curves are $\alpha = 0.69\pm$0.05 for reverse shock emission and
$t_b = 6.5\pm$0.2 day, $\alpha_1 = 0.99\pm$0.05 and $\alpha_2 =  2.0\pm$0.2 
for the OA. These parameters are improved further in the next
section by fitting the multi-wavelength observations with the standard
fireball model of GRBs. 

\begin{figure}
\begin{center}
\epsfig{file=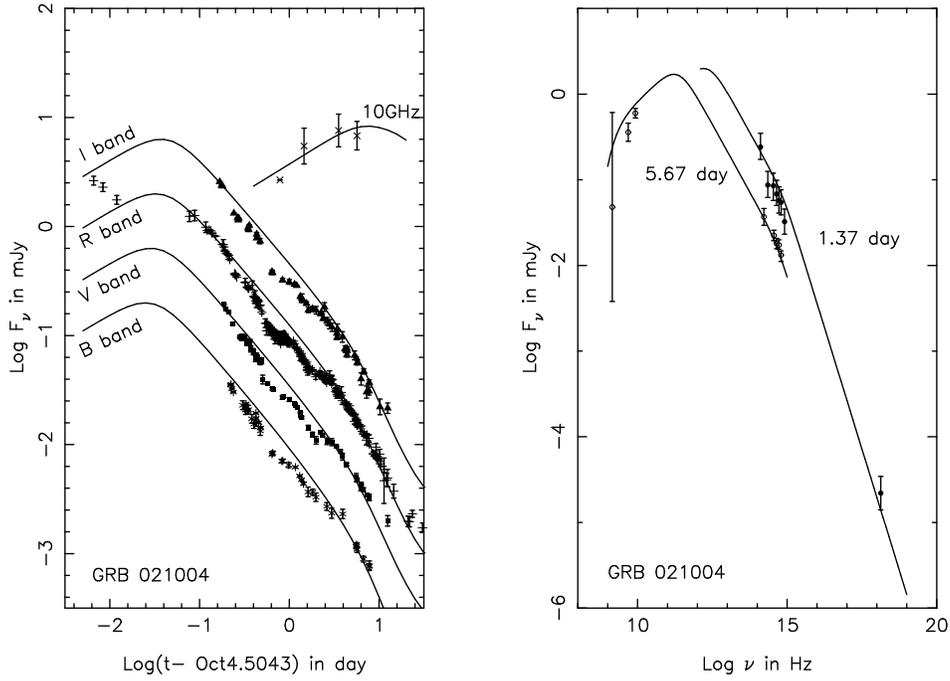,width=13cm}
\end{center}
\caption[x]{Multi-band observed light curves (left panel) and broadband spectrum 
(right panel) of the GRB 021004 OA are compared with model predictions shown as
solid curves. A total extinction of $E(B-V)$=0.20 mag has been used. For clarity 
of display, in the left panel the $B$,$V$,$I$ and radio light curves are shifted 
vertically by
$-1.0, -0.5, +0.5$ and $+1.0$ respectively in logarithmic scale. The radio light
curve at 10~GHz is constructed, by extrapolation with expected spectral slope, 
from measurements reported at 22.5~GHz, 15~GHz and 8.46~GHz at different epochs 
(Frail \& Berger 2002, Pooley et al. 2002, Berger et al. 2002). The frequency of 
10~GHz was chosen to correspond with a similar plot presented by Kobayashi \& 
Zhang (2003). The model uses $\nu_a = 2.1$~GHz, $\nu_m = 2.5\times10^{14}$~Hz 
and $\nu_c = 3.3 \times10^{16}$~Hz at $t=0.06$~day; a jet break time $t_b = 6.7$
day and an electron energy distribution index $p=2.27$. The model includes the
host galaxy contribution estimated in section 3.2. In the right panel an 
expected spectrum with $p = 2.27$ is shown with the Chandra HETG measurement 
(Sako \& Harrison 2002a) and optical and near-IR observations at the same epoch, 
$\sim 1.37$~day. A similar spectrum is also shown at $t=5.67$~day, the epoch of 
the cm-wave radio observations reported by Berger et al. (2002). Where necessary,
fluxes measured at optical, near-IR and radio wavelengths were interpolated 
between adjacent data points at one wavelength in order to determine a 
contemporaneous value with another wavelength.} 
\end{figure}

\section{Modelling of the GRB 021004 afterglow}

We attempt modelling the behaviour of GRB 021004 OA along the lines 
of standard GRB model proposed by Kobayashi \& Zhang (2003).  

The 7-400~keV gamma-ray fluence (Lamb et al. 2002) implies an 
isotropic-equivalent energy of $4.6\times 10^{52}$~erg emitted in the burst 
radiation, for $H_0=65$~km/s/Mpc in a $\Omega_{\rm m}=0.3$, $\Omega_{\rm 
\Lambda}=0.7$ cosmological model.  The corresponding comoving passband of
$23-1330$~keV contains the bulk of the emitted energy in most GRBs, and
a k-corrected estimate of Bolometeric energy is unlikely to exceed this
by more than $\sim 50\%$ (Bloom et al 2001).  Assuming a similar amount of 
energy to remain in the fireball to power the afterglow (Piran et al. 2001), 
one finds that for a typical $\epsilon_e\sim 0.1$ and $\epsilon_B \sim 10^{-2}$ 
(Panaitescu \& Kumar 2001) the frequency $\nu_m$ of maximum radiation 
in the afterglow spectrum should lie close to the optical band at $\sim 0.1$~day 
after the burst (Sari, Piran \& Narayan 1998; Kobayashi \& Zhang 2003). 
The brightening of GRB 021004 OA optical lightcurve at $\sim 0.1$~day, relative
to the extrapolated early decay, could therefore be attributed to 
the passage of $\nu_m$ through the optical band. The three early $R$-band 
observations (Fox 2002) which show a power-law decay of $\alpha = 0.69\pm$0.05
before the brightening can then be understood in terms of a decaying prompt 
emission from the reverse shock (Kobayashi \& Zhang 2003). We exclude 
this early emission from our further discussion and restrict ourselves to the 
properties of the forward shock emission.

In Fig. 3 we compare the predictions of a standard afterglow model with electron
energy distribution power-law index $p$=2.27 and a jet-break time $t_b$ of 6.7 
days. For this, the observed magnitudes/fluxes have been corrected for standard
Galactic extinction law given by Mathis (1990) and the effective wavelength
and normalization by Fukugita et al. (1995) for $U,B,V,R,I$ and by Bessell \&  
Brett (1988) for near-IR have been used. The fluxes thus derived are accurate 
to about 10\% in optical and about 25\% in near-IR.
For the model parameters mentioned above, relative normalization of the light 
curves in different optical passbands become consistent with the data if the total
extinction is $E(B-V)=0.20$~mag, which has been used in our analysis.
The Galactic extinction in this direction is estimated to be 
$E(B-V) = 0.06$~mag from the smoothed reddening map provided by 
Schlegel, Finkbeiner \& Davis (1998). The additional extinction may then be 
attributed to small scale fluctuations in the distribution of dust in our 
galaxy, or may, in part, originate even in the host galaxy of the GRB. 
The spectral slope deduced in section~2.2 from the HCT low resolution spectrum 
is also consistent with $p=2.27$ once the above total extinction 
is taken into account. The flux decay constants derived in the last
section are also consistent with the parameters used in the model.

Given the above model parameters we find that the broadband behaviour of the OA is 
well explained. However, such a model cannot reproduce short term variations, as 
seen during the interval $\sim 0.5-2$~days in the optical light curves (see Fig. 2). 
The reason for these short term variations could lie in density variations in the 
circum-burst medium, as conjectured by Lazzati et al. (2002) and 
Nakar et al. (2003).
Fig.~3 shows the broadband spectrum including $X-$ray, optical-near IR 
and radio observations and the model predictions. The cooling break $\nu_c$ is 
located between the optical and $X-$ray bands. The spectrum observed within the 
$X-$ray band at $\sim 1.37$~day (Sako \& Harrison 2002a) is $F_{\nu} \propto
\nu^{-1.1\pm0.1}$, and the decay rate is $t^{-1.0\pm0.2}$. Both are consistent 
with $p=2.27$ and $\nu > \nu_{c}$, at a time before the jet break. 
Our model predictions are also in good agreement with the Chandra observation 
at $t\sim 52$~day (Sako \& Harrison 2002b).   The presence of a jet break between 
the two observations results in the apparent temporal decay slope of $\sim 1.7$ 
reported by Sako \& Harrison (2002b).

The 1.4~GHz to 8.5~GHz radio spectrum at 5.67~days (Berger et al. 2002) is well 
fit by the model, assuming a self-absorption frequency $\nu_a$ near 2~GHz. The 
same assumptions lead to a good explanation of the cm-wave radio light curve. 
Fig.~3 shows the derived light curve at 10~GHz from measurements reported at 
nearby frequencies (Frail \& Berger 2002, Pooley et al. 2002, Berger et al. 2002) 
and the model prediction. However, this model is unable to 
reproduce the 85~GHz flux of 2.5~mJy observed at $\Delta t$ = 1.45~days (Bremer 
\& Castro-Tirado 2002). Although at the maximum of the spectrum ($\nu_m$) the
flux rises to 2.75~mJy in the model, at $\Delta t$ = 1.45 days, $\nu_m$ is 
located well above 85~GHz and the predicted flux is only $\sim 1$~mJy at 85~GHz.
It may well be that a part of the emission observed at 85~GHz comes from the 
host galaxy as in the case of GRB~010222 (Frail et al. 2002) which should be 
seen to remain visible after the afterglow fades away. 

\section{Discussions  and Conclusions }

We present the broad band $BVRI$ photometric and low-resolution spectroscopic
optical observations of the OA associated with GRB 021004 starting about 3 hour
after the burst. Our last photometric observations are at about $\Delta t$ = 
12 days. These observations in combination with the published multi-wavelength  
data have been used to study the flux decay and to derive parameters of the GRB 
and its afterglow. We have used secure photometric calibrations in the present
analyses. The optical observations obtained by Fox (2002) during the first 20 
minutes 
of the burst indicate that GRB 021004 is the second GRB OA after GRB 990123 from
which optical emission from the reverse shock has been observed (Galama et
al. 1999). The dense temporal $BVRI$ passband light curve indicates rapid light 
variations.  Such flux variations from a power-law decay have been reported only 
for GRB 000301c (Sagar et al. 
2000a, Masetti et al. 2000, Jensen et al. 2001, Garnavich, Loab \& Stanek 2000, 
Gaudi et al.  2001) and for GRB  011211 (Holland et al. 2002, Jakobsson et al. 
2003) so far. However, the amplitude of oscillation is maximum in the case of 
GRB 021004 OA being $\sim 0.5$~mag. The light curves show a steepening superposed 
on the achromatic, rapid variations which could be detected mainly due to the 
dense observations in $BVRI$ filters.
This indicates that in future the small telescopes, as large amount of observing 
time is available on them (cf. Sagar 2000), will  play an important
role in understanding the origin of such short term variability in the light
curves of GRBs during early times. The overall flux decay in observed light curves 
is well understood in terms of a jet model. The flux decay constants at early 
and late times derived from least square fits to the light curves are 
0.99$\pm$0.05 and 2.0$\pm$0.2 respectively. The value of the jet break time is 
about 7 day. The total extinction in the direction of the OA is
$E(B-V) = 0.2$ mag. The low-resolution spectrum corrected for this extinction
yields a spectral slope of $\beta = 0.6\pm$0.02. The photometric colour 
distributions determined in optical and near-IR regions for various epochs 
indicate that spectral index of the GRB 021004 afterglow has not changed 
significantly  during a period of about 15 days after the burst, while 
the flux decay slope has steepened from 1.0 to 2.0. GRB 021004 
thus becomes one more burst for which a clear achromatic break in the 
light curve is observed.  This is generally accepted as an evidence for 
collimation of the relativistic GRB ejecta in accordance with the prediction by 
recent theoretical models (M\'{e}sz\'{a}ros \& Rees 1999; Rhoads 1999; Sari
Piran \& Halpern 1999). 

Recent afterglow observations of GRBs show that a relativistic 
blast wave, in which the highly relativistic electrons radiate via synchrotron 
mechanism, provides a generally good description of the observed properties. 
In the case of GRB 021004 OA also, it appears that a standard fireball 
afterglow model, with a combination of emission from a forward and a reverse
shock can account for most of the overall behaviour of the afterglow. The observed 
fluxes, however, show unexplained fluctuations, falling significantly below 
model predictions in the range $\Delta t $ = 0.5--2~days. Density variations 
in the circum-burst medium is one possible explanation of this behaviour 
(Wang \& Loeb 2000, Lazzati et al. 2002, Nakar et al. 2003, Heyl and Perna 2003). 
While an alternative explanation in terms of micro-lensing 
(Garnavich, Loeb \& Stanek 2000) cannot be entirely ruled out, the multiple 
bumps seen in this light curve would not be natural in this model.
The observed jet break time of $\sim 7$~day, along with the burst fluence,
leads to an estimate of the jet opening angle of $\sim 7^{\circ}$ (for an 
assumed $\gamma-$ray efficiency $\eta_{\gamma}=0.2$ (Frail et al. 2001), and a 
circumburst density $n=0.3$~cm$^{-3}$ inferred from reverse-shock modelling by 
Kobayashi \& Zhang (2003), similar to the opening angles inferred in other jetted
afterglows (see, e.g.\ Panaitescu \& Kumar 2001). The inferred opening angle 
implies a total emitted gamma-ray energy of $E_{\gamma}\sim 3.5\times 10^{50}$ 
erg, close to the peak of the $E_{\gamma}$ distribution in GRBs as shown by 
Frail et al. (2001).  The modelling of the radio emission suggests that excess 
emission might have been detected at 85~GHz, possibly due to the emission from 
the host galaxy. If this turns out to be true, then the observed emission would 
indicate a strong star formation activity in the host galaxy. 
The multiple blue shifted H, C-IV and Si-IV absorption lines in the spectrum of
GRB021004 OA, with a velocity span of 3200 km/s, could be interpreted as 
a clumpy WC star wind environment (Mirabal et al. 2002b). However our modelling
indicates that the light curve after $\sim 0.1$~day is better explained by
a circumburst medium of nearly uniform density with small scale density 
variations rather than a $r^{-2}$ wind density profile.  This might imply a
variable mass loss rate in the wind (Heyl and Perna 2003) or that the 
circumburst medium is composed not of Wolf-Rayet wind but of expanding ejecta of 
a supernova preceding the burst (Salamanca et al 2002, Wang et al 2003).
In either case, these observations provide a strong support in favour of 
collapsar origin of this burst in particular, and of long duration GRBs in
general.

The peculiarity in the light curves of GRB 021004 could be noticed
mainly due to dense as well as multi-wavelength observations during early 
times. Such observations of recent GRBs have started 
revealing features which require explanation other than generally accepted 
so far indicating that there may be yet new surprises in GRB afterglows. 

\section*{Acknowledgements}
We thank an anonymous referee for comments which helped us improve the paper.
This research has made use of data obtained through the High Energy Astrophysics 
Science Archive Research Center Online Service, provided by the NASA/Goddard 
Space Flight Center. L. Resmi is supported by a CSIR research fellowship.

\label{lastpage}

\begin{thebibliography}{100}
\bibitem {} Anupama G.C., Sahu  D.K., Bhatt B.C., Prabhu T.P., GCN Observational
Report No. 1582 
\bibitem {} Barsukova E.A., Goranskij V.P., Bestin G.M., Plokhotnichenko V.L., Pozanenko A.S., 2002 GCN Observational Report No. 1606
\bibitem {} Berger E., Sari R., Frail D. A., et al., 2000, ApJ, 545, 56 
\bibitem {} Berger E., Frail D. A., Kulkarni S.R., 2002, GCN Observational Report No. 1612, 1613 
\bibitem {} Bersier D., et al., 2003, ApJL, in press (astro-ph/0211130)
\bibitem {} Bessell M.S., Brett J.M., 1988, PASP, 100, 1134 
\bibitem {} Bloom J.S., Frail D.A., Sari R., 2001, ApJ, 121, 2879
\bibitem {} Bremer M., Castro-Tirado A.J., 2002, GCN Observational Report No. 1590 
\bibitem {} Castro-Tirado A.J., Perez E., Gorasabel J., et al., 2002 GCN Observational Report No. 1635 
\bibitem {} Chornock R.,  Filippenko A.V., 2002, GCN Observational Report No. 1605 
\bibitem {} Covino S., Ghisellini G., Malesani D., et al., 2002, GCN
Observational Report Nos. 1595, 1622 
\bibitem {} Cowsic R., Prabhu T. P., Anupama G. C. et al., 2001, BASI, 29, 157
\bibitem {} Di Paola A., Boattini A., Del Principe M., Konstantinova T., Larionor V., Antonelli L., 2002, GCN Observational Report No. 1616 
\bibitem {} Djorgovski S.G., Barth A., Price P., et al., GCN Observational Report No. 1620 
\bibitem {} Doty J., Grew G.,  Jernigan J.G., et al. 2002, GCN Observational Report No. 1568 
\bibitem {} Eracleous M., Schaeter B.E. Moder J., Wheeler G., 2002, GCN Observational Report No. 1579 
\bibitem {} Filippenko A.V., 1982, PASP, 94, 715                
\bibitem {} Fox D.W., 2002, GCN Observational Report No. 1564 
\bibitem {} Fox D.W., Barth A.J., Soderberg A.M., et al., 2002, GCN Observational Report No. 1569 
\bibitem {} Frail et al. 2001, ApJ, 562, L55
\bibitem {} Frail D.A., Berger E., 2002, GCN Observational Report No. 1574 
\bibitem {} Frail et al. 2002, ApJ, 565, 829
\bibitem {} Fukugita  M., Shimasaku K., Ichikawa T., 1995, PASP, 107, 945 
\bibitem {} Galama T.J. et al., 1999, Nature, 398, 394 
\bibitem {} Garnavich P.M., Loeb, A.\& Stanek K. J., 2000, ApJ, 544, L11 
\bibitem {} Garnavich P., Quinn J., 2002, GCN Observational Report No. 1661 
\bibitem {} Gaudi B.S., Granot J., Loeb, A. 2001, ApJ, 561, 178  
\bibitem {} Halpern J.P., Armstrong E.K., Espaillat C.C., Kemp J., 2002a, GCN Observational Report No. 1578
\bibitem {} Halpern J.P. Mirabal N., Armstrong E.K., Espaillat C.C.., Kemp J., 
    2002b, GCN Observational Report No. 1593 
\bibitem {} Harrison et al., 1999, ApJ, 523, L121
\bibitem {} Harrison F. A., Yost S. A., Sari R. et al., 2001, 559, 123
\bibitem {} Henden A., 2002, GCN Observational Report Nos. 1583, 1630 
\bibitem {} Henden A. \& Levine S., 2002,  GCN Observational Report No. 1592 
\bibitem {} Heyl J.S. \& Perna R., 2003, ApJL, in press (astro-ph/0211256)
\bibitem {} Holland S. T., Soszynski I., Gladders M., et al., 2002, AJ, 124, 639 
\bibitem {} Holland S. T. et al., 2003, AJ, submitted (astro-ph/0211094)
\bibitem {} Jakobsson P., Hjorth J.,  Fynbo J.U. et al., 2003, submitted to A\&A
\bibitem {} Jensen B.L., Fynbo J.U., Gorosabel J. et al., 2001, A\&A, 370, 909 
\bibitem {} Kobayashi S., Zhang B., 2003, ApJ, 582, L75
\bibitem {} Lamb D., Ricker G., Atteia J-L., et al., GCN Observational Report No. 1600 
\bibitem {} Landolt, A.R., 1992, AJ, 104, 340 
\bibitem {} Lazzati D.,  Rossi E., Covino S., Ghisellino, G., Malcsani D., 2002, A\&A, 396, L5 
\bibitem {} Malesani D., Covino S., Ghisellini G., et al., 2002a, GCN Observational Report No. 1607 
\bibitem {} Malesani D., Stefanon M., Covino S., et al., 2002b, GCN Observational Report No. 1645 
\bibitem {} Masetti N. et al., 2001, A\&A, 374, 482 
\bibitem {} Masetti N., Pizzichini G., Bartolini C., et al., 2002, GCN Observational Report No. 1603 
\bibitem {} Mathis J.S., 1990, ARAA, 28, 37 
\bibitem {} Matheson T., Garnavich P.M., Foltz G. C., et al., 2003, ApJ, 582, L5
\bibitem {} Matsumoto K., Kawabata T., Ayani K., Urata Y., Yamaoka H., Kawai N., 
2002, GCN Observational Report No. 1594 
\bibitem {} M\'{e}sz\'{a}ros P., Rees M. J., 1999, MNRAS, 306, L39 
\bibitem {} Mirabal N., Armstrong E.K., Halpern J.P.,  Kemp J., 2002a, GCN Observational Report No. 1602 
\bibitem {} Mirabal N., Halpern J.P., Chornock R., Filippenko A.V., 2002b, GCN Observational Report No. 1618
\bibitem {} M$\o$ller, P., Fynbo, J.P.U., Hjorth J. et al., 2002, A\&A, 396, L21  
\bibitem {} Nakar, E., Piran,T., Granot,J., 2002, Submitted to New Astronomy (astro-ph/0210631)
\bibitem {} Oksanen A., Aho M., Rivich K., Rivich K., West D., Durig D., 2002, GCN Observational Report No. 1591 
\bibitem {} Panaitescu A., Kumar P., 2001, ApJ, 560, L49
\bibitem {} Panaitescu A., Kumar P., 2002, ApJ, 571, 779 
\bibitem {} Pandey S.B., Sagar R., Mohan V., Pandey A.K, Bhattacharya D., \& Castro-Tirado A.J., 2001, BASI, 29, 459 
\bibitem {} Piran T., Kumar P., Panaitescu A., Piro L. 2001, ApJ, 560, L167
\bibitem {} Pooley G., 2002, GCN Observational Report Nos. 1575, 1588, 1604 
\bibitem {} Rhoads J.E., 1999, ApJ, 525, 737 
\bibitem {} Rhoads J. E. \& Fruchter A., 2001, ApJ, 546, 117
\bibitem {} Rhoads J., Burud J., Freuchter A., 2002, GCN Observational Report No. 1601 
\bibitem {} Rol, E. et al., 2002, GCN Observational Report No. 1596 
\bibitem {} Sagar R., 2000, Current Science, 78, 1076
\bibitem {} Sagar R., 2001, BASI, 29, 215
\bibitem {} Sagar R., 2002, BASI, 30, 237
\bibitem {} Sagar R., Mohan V., Pandey S.B., Pandey A.K., Stalin C.S., Castro-Tirado A.J., 2000, BASI, 28, 499 
\bibitem {} Sagar R., Pandey S.B., Mohan V., Bhattacharya D., Castro-Tirado A.J., 2001a, BASI, 29, 1,
\bibitem{} Sagar R., Stalin C. S., Bhattacharya D., Pandey S. B., Mohan V., Castro Tirado A. J., 
Pramesh Rao A., Trushkin S. A., Nizhelskij N. A., Bremer M. and Castro Cer\'{o}n J.  M., 2001b, 
BASI, 29, 91
\bibitem {} Sahu D.K., Bhatt B.C., Anupama G.C., Prabhu T.P., GCN Observational Report No. 1587 
\bibitem {} Salamanca I., Rol E., Wijers R., Ellison S., Kaper L., Tanvir N., 2002, GCN Observational Report No. 1611 
\bibitem {} Sako M., Harrison F.A., 2002a, GCN Observational Report No. 1624 
\bibitem {} Sako M., Harrison F.A., 2002b, GCN Observational Report No. 1716 
\bibitem {} Sari R., Piran T., Halpern J. P., 1999, ApJ, 519, L17 
\bibitem {} Sari R., Piran T., Narayan R., 1998, ApJ, 497, L17 
\bibitem {} Savaglio S., Fiore F., Israel F. et al., 2002, GCN Observational Report No. 1633 
\bibitem {} Schlegel D.J., Finkbeiner D.P., Davis M., 1998, ApJ, 500, 525 
\bibitem {} Shirasaki C., Graziani M., Matsuoka M., et al., 2002, GCN Observational Report No. 1565 
\bibitem {} Stanek K. Z. et al., 1999, ApJ, 522, L39 
\bibitem {} Stanek K. Z. et al., 2001, ApJ, 563, 592 
\bibitem {} Stefanon M., Covino S., Malesani D. et al., 2002, GCN Observational Report No. 1623 
\bibitem {} Wang L., Baade D., Hoeflich P., Wheeler C. J., 2002, GCN Observational Report No. 1672 
\bibitem {} Wang L., Baade D., Hoeflich P., Wheeler C. J., 2003, Submitted to ApJL (astro-ph/0301266) 
\bibitem {} Wang X., Loeb A., 2000, ApJ, 535, 788
\bibitem {} Weidinger M., Egholm M. P., Fynbo J.P.U., et al., 2002, GCN
Observational Report No. 1573
\bibitem {} Williams G., Lindsay K., Milne P.,  2002, GCN Observational Report No. 1652 
\bibitem {} Winn J., Bersier D., Stanek K.Z., Gernavich P., Walker A., 2002, GCN Observational Report No. 1576 
\bibitem {} Wood-Vasey W.M., Aldering G., Lee B.C. et al., GCN Observational Report No. 1572 
\bibitem {} Zharikov S., Vazquez R., Benitez G., del Rio S., 2002, GCN Observational Report No. 1577 
\end{thebibliography}
\end{document}